\documentstyle[prl,aps,epsf,twocolumn]{revtex}
\begin{document}
\draft
\title
{\bf Atomic Scale  Sliding and Rolling of Carbon Nanotubes}
\author{A.~Buldum$^{\ast}$ and Jian Ping ~Lu$^{\dagger}$} 
\address{
Department of Physics and Astronomy, The University of North Carolina 
at Chapel Hill, Chapel Hill, NC, 27599}
\date{\today}
\maketitle
\begin{abstract}
A carbon nanotube is an ideal object for understanding 
the atomic scale aspects of interface interaction and friction. 
Using molecular statics and dynamics methods  
different types of motion of nanotubes on a graphite surface are 
investigated. We found that
each nanotube has unique equilibrium orientations with sharp 
potential energy minima. This leads to atomic scale locking 
of the nanotube.  
 The effective contact area and the total interaction energy 
scale with the square root of the radius.
Sliding and rolling of nanotubes have different characters. The potential
energy barriers for sliding nanotubes  are higher than that 
for perfect rolling. When the nanotube is pushed, we observe a combination 
of atomic scale spinning and sliding motion. The  result is
rolling with the friction force  comparable to sliding.
\end{abstract}
\narrowtext
 Although the fundamental aspects of friction 
have been studied for more then centuries, 
our knowledge about its microscopic aspects is very limited\cite{1}. 
The invention 
of atomic force microscope\cite{2} (AFM) and its application in 
measurements of atomic scale  friction\cite{3} (friction 
force microscope - FFM) have made a great impact on the studies of friction.
A carbon nanotube is a stable nanoobject having  cylindrical shape\cite{4}, thus 
ideal for understanding  atomic scale friction.
M. Falvo et. al. showed that it is possible to  slide, rotate and roll 
carbon nanotubes on a graphite surface\cite{5}. 
They demonstrated that a nanotube has preferred orientations 
on the graphite surface and prefer
rolling than sliding when it is in atomic scale registry with the surface. 
In this study, we carried out molecular statics, dynamics calculations 
and studies of stick-slip motion for a variety of nanotubes. 
We found that: (i) A nanotube has sharp potential energy minima 
leading to orientational locking. The  locking angles are 
directly related to the chiral angle.(ii) Sliding and rolling of nanotubes have 
different characters.  The energy barriers for sliding 
are higher then the barriers for perfect rolling.  
(iii) The effective contact area and total interaction energy
scale with the square root of the radius.
(iv) A combination of sliding and spinning motion is observed
 when the tube is pushed. The net result is rolling with the friction force
comparable to the corresponding force for sliding. \\
\\
The character of interaction  between the moving object ( atom, 
molecule or any nanoparticle ) and the underlying surface defines the 
motion\cite{6}. The interaction energy may consist of short-range,
attractive interaction  energy due to chemical bonding;
short-range repulsive energy and  long-range, attractive van der Waals
energy. The interaction between a carbon nanotube and 
a graphite surface is similar to that between two 
graphite planes which is  weak and van der Waals in origin. 
To investigate the overall behavior of the motion of a carbon nanotube
on a graphite surface, we represent the interaction between the tube and
the graphite surface atoms by an empirical potential of Lennard-Jones type\cite{7}
which was used extensively to study solid $C_{60}$ and nanotube\cite{8}.
Recent theoretical calculations\cite{9} showed that multiwall nanotubes on a 
graphite surface are not deformed significantly. 
The atomic scale motion is determined mostly by the interaction of the outmost
layer of the nanotube with the surface. 
 In this work we studied rigid single wall nanotubes with different
chiralities and radii. 
\begin{figure}
\vskip -0.8in
\centerline{
\epsfxsize=3.0in \epsfbox{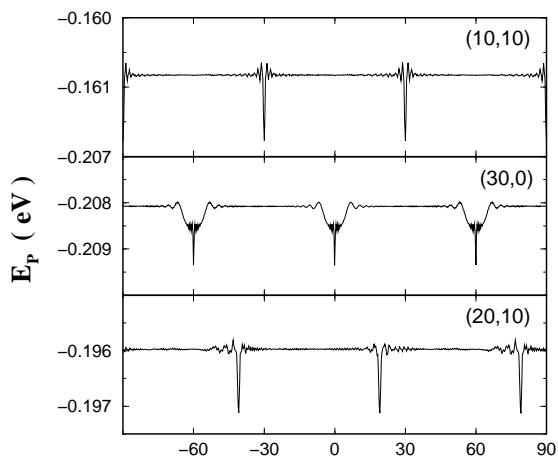}
}
\vskip -0.5in
\caption{The interaction energy
as a function of rotation angle between the nanotube axis and 
the graphite lattice
for (10,10), (30,0) and (20,10) nanotubes.
Each nanotube has unique minimum energy orientations 
repeating  in every 60 degrees. }
%\label{transistor}
\end{figure}
\noindent
The energy barriers related to the motion of nanotubes can be conveniently 
analyzed by calculating the variation of the potential energy, $E_{P}$, and
corresponding force during the motion. 
Four different
types of motion are considered: 
spinning, rotating, sliding and rolling. 
During each step of motion the height of the tube is optimized. 
We first spin and rotate the nanotubes in order to find
the equilibrium positions.  
Fig. 1 shows the interaction energy as a function of the rotation angle 
between the tube axis and the graphite lattice (All the data given in this
study is for per \AA\,\,\, length of the nanotubes).   
Each nanotube has unique 
equilibrium orientations repeating in every 60$^o$, reflecting the
lattice symmetry of the graphite. 
The variation of energy near 
the minimum is very sharp which causes atomic scale locking of the nanotube.
Locking angles are different for different nanotube and they are the 
direct measure of the chiral angle. 
This provides a novel method for measuring the chiralities of 
carbon nanotubes. Other important point in Fig. 1 is  that the 
energy variation between two consecutive energy minima is 
very small (except near the minima) . Thus  
 the force  needed to rotate a nanotube is very small 
when the tube is out-of-registry.    
These results 
are in good agreement with the recent experiments by M. Falvo et. al. \\
\begin{figure}
%\vskip -0.5in
\centerline{
\epsfxsize=2.5in \epsfbox{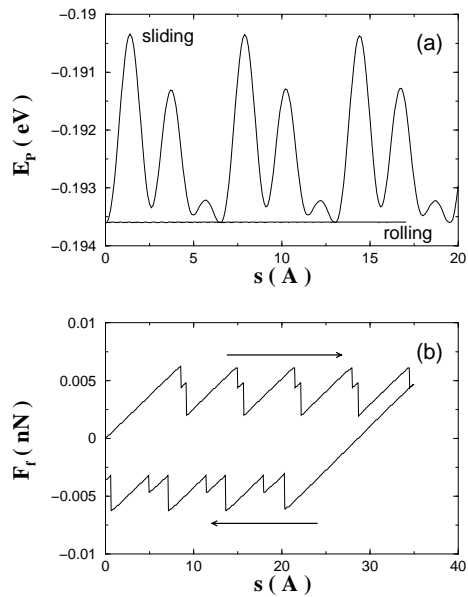}
}
\vskip -0.2in
\caption{(a) The variation of the interaction
energy,$E_{P}$  of a (20,10) nanotube
as a function of sliding and rolling distances. 
(b) The  friction 
force in stick-slip motion of the (20,10) nanotube 
when attached to an AFM tip with spring constant $\sim 8.0\times10^{-3}$ N/m.}
\end{figure}
Next, we studied sliding and rolling of carbon nanotubes. 
In Fig. 2(a) the  variation of the total interaction
energy, $E_{P}(s)$ of a (20,10) nanotube 
as a function of sliding distance, $s$ is shown.
The energy variation is very small for rolling since 
perfect atomic scale registry is always maintained in the contact region. 
On the other hand in sliding motion, all the atoms in the contact 
region move simultaneously out-of-registry to higher energy positions  
and contribute to the  energy barrier of motion. 
When the nanotube is attached to an AFM tip,
stick-slip motion occurs. 
The tube first sticks and then
slips suddenly (slides or rolls or both) when the force exerted by the 
tip is sufficiently large\cite{10}. 
The friction force in stick-slip motion of 
the  nanotube  when it is attached to an AFM tip ( with string constant 
$\sim 8.0\times10^{-3}$ N/m )
is shown in Fig. 2(b).  
The area in the hysteresis curve gives us the amount of energy dissipated 
during the stick-slip motion. \\
\\
To investigate the tube dependence, we performed calculations with 
tubes in different chiralities or radii.   
Fig. 3(a) shows the interaction energy as 
a function of the nanotube radius, $R$. 
We found  that the effective contact area and the  interaction
energy scale with the  square root of the radius of the nanotube. 
  The interaction energy is independent of chirality.
In-registry sliding force (when the tube is in a minimum energy
orientation) is also scale with 
the square root of the radius.
However, the sliding force is 
different for nanotubes with different chiralities and the same radii.
Fig. 3(b) shows the force for in-registry sliding and rolling.
 For a typical nanotube 
( radius $\sim 13$ nm, length $\sim$ 600 nm ), the sliding force value
is estimated as $\sim 87$ nN for an armchair tube and $\sim 43$ nN for a
zigzag tube in good agreement with 
the friction force values measured in the experiments\cite{5}
( $\sim 50$ nN ). 
\\
\begin{figure}
\vskip -0.7in
\centerline{
\epsfxsize=2.9in \epsfbox{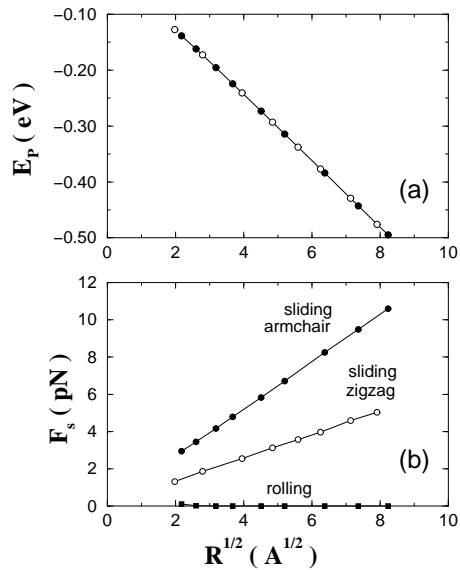}
}
\vskip -0.2in
\caption{(a) The interaction energy as 
a function of square root of the nanotube
radius, $\sqrt{R}$.
The filled circles  correspond to armchair tubes and the hollow circles
correspond to zigzag tubes with different radii. 
 (b)  The corresponding force for in-registry 
sliding and rolling of the nanotubes as a function of $\sqrt{R}$. }
\end{figure}
\noindent
 The static calculations we discussed 
gives  insight of  ideal sliding and rolling.  However, the
dynamical behavior is crucial for the competition between  sliding and
rolling in the course of motion. In our model, the graphite substrate 
and the nanotube are considered to be rigid but the nanotube as a 
whole is able to move having translational and spinning degrees of freedom. 
Constant lateral force was applied to the nanotube for a short period of time 
( 50 ps ) and then 
the motion of the nanotube was analyzed. In the same way, 
 we applied constant torque or combinations of torque and lateral force 
to the nanotube. Total energy of the system was kept constant. \\
\\
\begin{figure}
\vskip -0.1in
\centerline{
\epsfxsize=2.8in \epsfbox{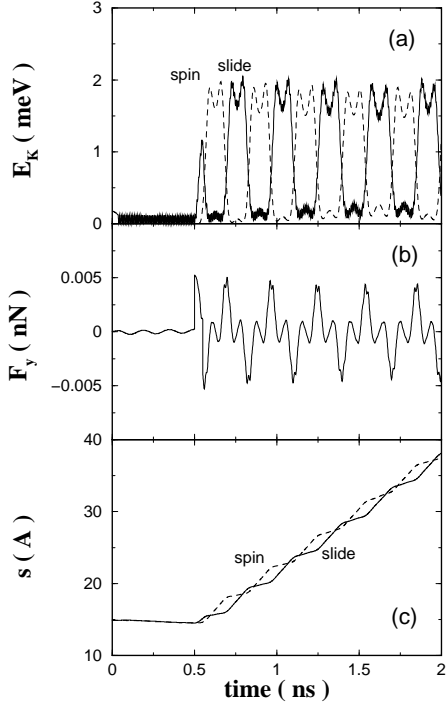}
}
\vskip -0.2in
\caption{(a) After initial push, sliding
and spinning components of the kinetic energy of the (10,10) nanotube as 
functions of time are represented by the solid and 
dashed lines respectively. Notice the switching between spinning and 
sliding motions  in the atomic scale. (b) The total force acting on the tube 
in the direction of motion
(c) The sliding and spinning distances (angle multiplied by $R$) 
as functions of time. }
\end{figure}
\noindent
After constant lateral force is applied on the nanotube, 
slide-spin motions in the atomic scale are observed. 
When the atoms in the contact region are in atomic scale registry 
it is easier for the nanotube to slide. Then the nanotube atoms 
move from in-registry to out-of-registry positions
and it is easier for 
the nanotube to spin. By spinning the nanotube decreases its potential energy and
the atoms recover the in-registry positions. 
 The switching of the tube motion 
between spinning and sliding can be clearly seen in the sliding
and the spinning component of the kinetic energy shown in Fig. 4(a). 
The switching between spinning and 
sliding is in the atomic scale and directly related to the corrugation of
interaction energy.  
The total force acting on the tube in the direction of motion is shown in Fig 4(b).
The maximum value of the force is comparable to the force required to slide the
nanotube (see Fig. 3(b)). Sliding and spinning distances ( angle multiplied by $R$)
 as  functions of time are shown 
in Fig. 4(c). Ideal rolling would be a perfect overlap of sliding and
spinning distances
meanwhile  slide-spin
motion gives oscillations and the net 
result is equivalent to rolling. \\
\\
For a better understanding of the  slide-spin  motion,
we plot the interaction  energy as a function of 
sliding distance  and spinning angle  in Fig. 5(a).
The trajectory for ideal rolling is a line 
at the bottom of the valley like regions. 
However, a nanotube performing 
 slide-spin motion follows an oscillating path in these valleys (see Fig. 5(b)) with 
an amplitude of oscillation depending on the initial kinetic energy. 
\begin{figure}
\vskip -0.2in
\centerline{
\epsfxsize=2.6in \epsfbox{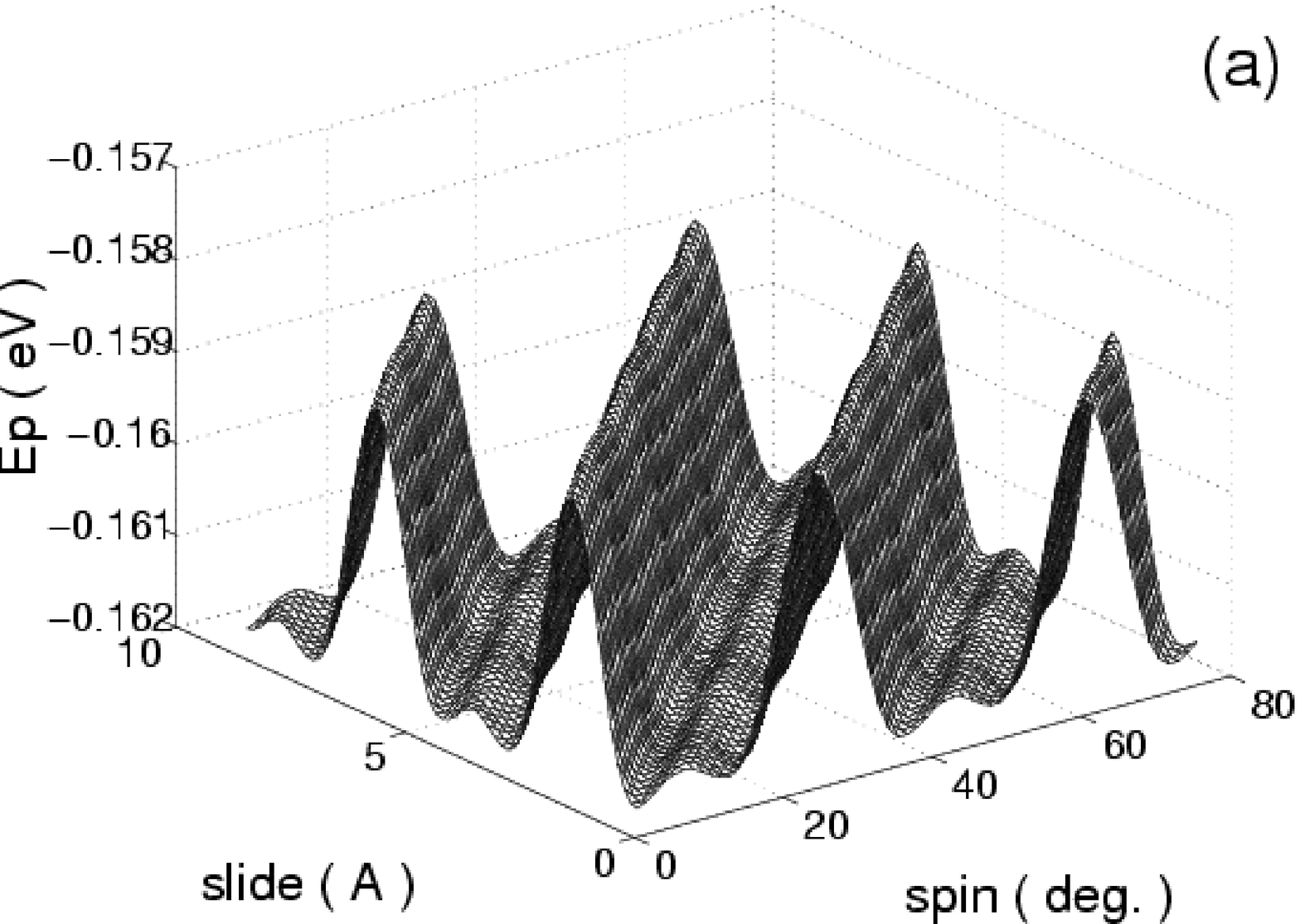}
}
\vskip -0.2in
\centerline{
\epsfxsize=2.6in \epsfbox{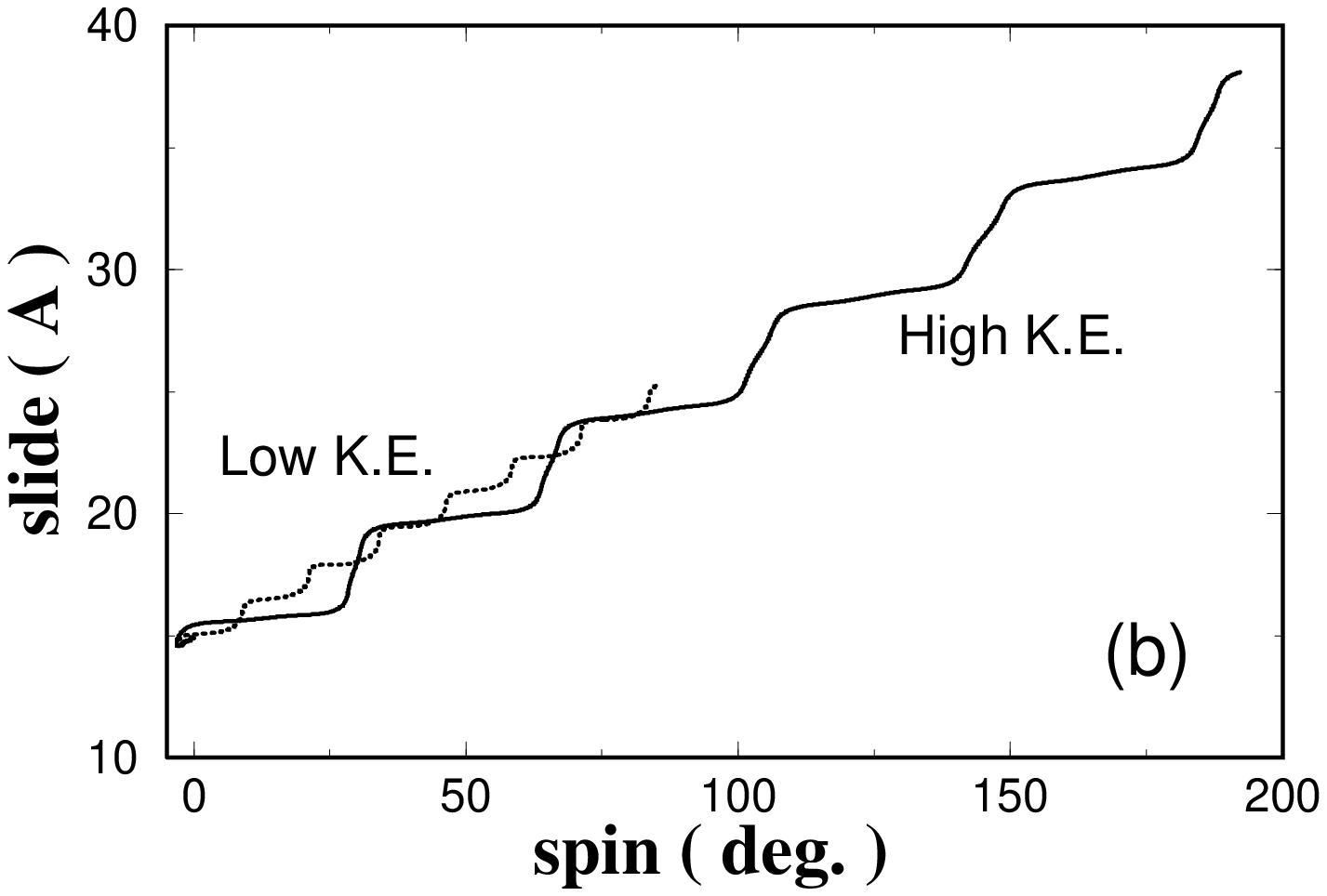}
}
\vskip -1.5in
\caption{ (a) The variation  of the interaction  energy as a function of
sliding distance and spinning angle. (b) The trajectories correspond to the 
slide-spin motion of a nanotube having lower and
higher total kinetic energies on the plane defined in (a)}
\end{figure}
\noindent
When the system is coupled to
a heat bath or energy dissipation due to friction is considered,
there are changes in the slide-spin motion. 
 We modeled the dissipation
by additional velocity dependent 
forces\cite{11} on the
atoms close to the contact region. The results are presented in
Fig. 6. 
Without energy dissipation 
the tube oscillates in the valley like regions of potential
energy surface (see Fig. 5(a)) in the slide-spin motion. 
When the energy dissipation is considered the total kinetic energy decreased
  and  there is more mixing between 
sliding and spinning.  Eventually, the  nanotube performs ideal rolling. 
If the nanotube has very high  
kinetic energy, it slides over many surface unit cells.  But due to energy dissipation
the tube's total kinetic energy decreases and  slide-spin
motion starts. Afterwards the motion is close to ideal rolling (As seen in Fig. 6(b)).
This atomic scale picture of rolling is very similar  to the rolling of
macroscopic objects. Recent molecular dynamics  simulations\cite{12} with a 
full relaxed system  by J. D. Schall and D. W. Brenner 
find similar conclusions. \\
\\ 
\begin{figure}
\vskip -0.2in
\centerline{
\epsfxsize=2.5in \epsfbox{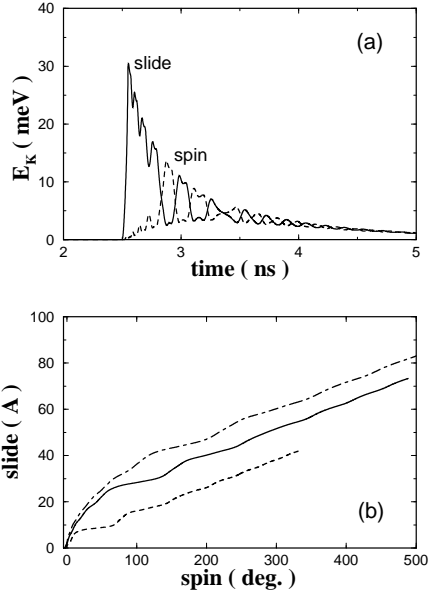}
}
%\vskip -0.2in
\caption{(a)Sliding and spinning component of the kinetic energy of the 
(10,10) nanotube as functions of time are represented by the solid  
and dashed lines respectively (b)  The nanotube's trajectories on the surface 
defined in fig. 5(a) for different initial kinetic energies. The trajectories from
higher to lower initial kinetic energies are represented by dotted-dashed, solid and
dashed lines, respectively. The kinetic energy values in (a) corresponds to the trajectory
plotted by solid line.}
\end{figure}
\noindent
To conclude, we investigated different types of motion of carbon nanotubes
on a graphite surface. 
Each nanotube has unique minimum energy orientations with respect 
to the surface structure. 
The variation of interaction  energy is
very sharp leading to orientational locking of the nanotubes. 
The locking angles are direct measure of the chiral angles 
 and this provides a novel method  for measuring
the nanotube chirality.
  We found that the effective contact area and the total interaction potential 
energy scale with square root of the radius of the nanotube. 
A combination of atomic scale spinning and sliding motion is observed when the 
nanotube is pushed. 
\begin{acknowledgements}
The authors are greatful to R. Superfine, M. R.  Falvo, J. Steele, 
D.W. Brenner, J.D. Schall and S. Washburn for many  stimulating discussions.
This work is supported by U. S. Office of Naval Research (N00014-98-1-0593).
\end{acknowledgements}
 
\vspace{0.5cm}
\noindent
$^{\ast}$ Email: buldum@physics.unc.edu \\
$^{\dagger}$ Email: jpl@physics.unc.edu

\end{document}